\documentclass[twocolumn,prd,nofootinbib,aps,prl,floats,floatfix,amsmath,amssymb,longbibliography,secnumarabic]{revtex4-1}
\pdfoutput=1
\usepackage[final]{graphicx}
\usepackage{hyperref}
\usepackage{amsmath}
\usepackage{bbm}
\usepackage{amsfonts}
\usepackage{amssymb}
\usepackage{latexsym}
\usepackage{graphicx}
\usepackage[english]{babel}
\usepackage{multirow}
\usepackage{float}
\usepackage{url}
\usepackage{slashed}
\usepackage{xcolor} 
\usepackage[utf8]{inputenc}

\def\sfrac#1#2{{\textstyle{#1\over #2}}}
\newcommand{\be}{\begin{equation}}
\newcommand{\ee}{\end{equation}}
\newcommand{\ba}{\begin{array}}
\newcommand{\ea}{\end{array}}
\newcommand{\bea}{\begin{eqnarray}}
\newcommand{\eea}{\end{eqnarray}}
\newcommand{\sss}{\scriptscriptstyle}

\newcommand{\nn}{\nonumber}
\renewcommand{\L}{{\sss L}}
\newcommand{\R}{{\sss R}}

\begin{document}

\title{Dark decay of the neutron}

\author{James M.\ Cline}
 \email{jcline@physics.mcgill.ca}
\author{Jonathan M.\ Cornell}
 \email{cornellj@physics.mcgill.ca}
\affiliation{McGill University, Department of Physics, 3600 University St.,
Montr\'eal, QC H3A2T8 Canada}
\begin{abstract}

New decay channels for the neutron into dark matter plus other
particles have been suggested for explaining a long-standing
discrepancy between the neutron lifetime measured from trapped 
neutrons versus those decaying in flight.  Many such scenarios are
already ruled out by their effects on neutron stars, and the decay
into dark matter plus photon has been experimentally excluded. Here we
explore the decay into a dark Dirac fermion $\chi$ and a dark photon $A'$,
which can be consistent with all constraints if $\chi$ is a
subdominant component of the dark matter.  Neutron star constraints 
are evaded if the dark photon mass to coupling ratio is
$m_{A'}/g' \lesssim (45-60)\,$MeV, depending upon the nuclear equation
of state.  $g'$ and the kinetic mixing between U(1)$'$ and electromagnetism are tightly constrained by direct and indirect dark matter detection, supernova constraints, and cosmological limits.

\end{abstract}
\maketitle

%--------------------------------------------------------------------------------------------------

\section{Introduction}   Recently a
long-standing experimental discrepancy in the lifetime of the
neutron has been highlighted: the decay rate of neutrons stored in a bottle \cite{Mampe:1993an, Serebrov:2004zf, Pichlmaier:2010zz,Steyerl:2012zz,Arzumanov:2015tea} seems to be 
larger than that determined by detecting the decay products from
neutrons in flight \cite{Byrne:1996zz,Yue:2013qrc}.  The difference in
the associated lifetimes is $\sim 8$\,s with a $4\sigma$ tension.  
Ref.\ \cite{Fornal:2018eol} pointed out that this could be explained
by a hitherto undetected extra decay channel into dark particles,
if the neutron mixes with fermionic dark matter (DM) $\chi$ whose mass is in a
narrow range $m_p-m_e < m_\chi < m_n$ consistent with proton
stability.  The simplest decay channel, $n\to\chi\gamma$, comes from the transition
magnetic moment interaction $\bar n\sigma_{\mu\nu}\chi F^{\mu\nu}$
induced by the $n$-$\chi$ mixing.  This has already been ruled out
by a null search for the monochromatic photon \cite{Tang:2018eln}.
(It was also argued that any such solution of the lifetime problem
would exacerbate tensions in discrepant determinations of the neutron
axial-vector coupling $g_A$ \cite{Czarnecki:2018okw}). 

Even if the new decay channel is completely invisible, for example
$n\to\chi\phi$ where $\phi$ is a dark scalar, neutron stars ostensibly
rule out the scenario
\cite{McKeen:2018xwc,Baym:2018ljz,Motta:2018rxp}, since $\chi$
typically has a much softer equation of state than nuclear matter at
high density.  The conversion of $n$ to $\chi$ in a neutron star (NS) then
leads to the maximum possible mass of the star being well below the
largest values observed (near $2 M_\odot$).   Ref.\ 
\cite{McKeen:2018xwc} pointed out that this constraint can be evaded
if $\chi$ has additional pressure from the repulsive 
self-interactions that would come from dark photon ($A'$) exchange, if
the mass-to-coupling ratio $m_{A'}/g'$ is sufficiently large, though
the minimum required value was not determined.  One of
our goals is to find this value.

This suggests that a consistent picture can be made if the new dark
decay channel is $n\to \chi A'$.  
Astrophysical and cosmological considerations substantially constrain 
DM in the mass range required to explain the neutron lifetime anomaly.
However, $\chi$ could still be a subdominant
component of the total dark matter.

In the following, we start by deriving constraints on our scenario
that depend only upon the low-energy particle content at the GeV scale
or below, in Section \ref{mod_indep_sec}.  This is followed by
analysis of the effects upon neutron stars, Sec.\ \ref{NS_sec}, where
an upper bound on $m_{A'}/g'$ is derived.  We introduce the
microscopic renormalizable model in Sec.\ \ref{UV_sec} and discuss
the implications for cosmology in Sec.\ \ref{cosmo_sec}.  The
constraints from direct and indirect detection are derived in Sections
\ref{DD_sect} and \ref{ind_det_sec}.  Various constraints on the kinetic mixing 
parameter between the U(1)$'$ and electromagnetic $U(1)$ field
strengths are compiled in Sec.\ \ref{eps_sec}, followed by our
conclusions, Sec.\ \ref{conc_sec}.

\section{\bf Model-independent analysis}
\label{mod_indep_sec}  Before presenting a more
complete particle physics model, we start with the low-energy
effective Lagrangian that contains almost everything needed for the
phenomenology.  It has mixing of the Dirac $\chi$ with the
right-handed component of the neutron (this restriction does come from
UV considerations, to be explained later) and the usual terms for
$\chi$ interacting with a dark photon $A'$:
\bea
	{\cal L}_{\rm eff} &=& 
	\bar\chi(i\slashed{D} - m_\chi)\chi + \bar n \left (i \slashed{\partial} - m_n + \mu_n \sigma^{\mu \nu}F_{\mu \nu} \right) n \nn \\
	&-&\sfrac14 F_{\mu\nu}' F'^{\mu\nu} -\sfrac12 m_{A'}^2 A'^\mu A'_\mu\nn\\
	&-&\delta m\, \bar n_\R\chi_\L +{\rm h.c.}  -{\epsilon\over 2}
	F_{\mu\nu}F'^{\mu\nu}  
\label{Leff}
\eea
where $D_\mu = \partial_\mu -ig'A'_\mu$ and $\mu_n$ is the neutron magnetic dipole moment.  Kinetic mixing with the photon
is included so that $A'$ will eventually decay to photons through
a loop diagram, or electrons if $m_{A'}> 2 m_e$.  We assume that
$\chi$ is a Dirac particle, since UV models in which $\chi$ is
Majorana generically lead to dinucleon decays such as $^{16}$O$(pp)\to$
$^{14}$C$\pi^+\pi^-$ \cite{Aitken:2017wie}.

Diagonalization of the $n$-$\chi$ mass matrix requires 
rotations of $(n_R,\chi_R)$
and $(n_L,\chi_L)$ respectively by angles $\theta_R$ and $\theta_L$
with
\be
	\theta_R \cong {m_n\delta m\over m_n^2-m_\chi^2},\quad
	\theta_L \cong {m_\chi\delta m\over m_n^2-m_\chi^2}
\label{mixing}
\ee
Since $m_\chi$ is close to $m_n$, these angles are roughly equal,
$\theta = \theta_{R}\cong\theta_L\cong \delta m/(2(m_n-m_\chi))$.  Going to the
mass basis, the $\bar\chi\slashed{A'}\chi$ interaction then leads to
an off-diagonal term,
\be
	-g'\theta\, \bar n\slashed{A'}\chi + {\rm h.c.}
\ee
that leads to the dark decay $n\to \chi A'$ assuming that $m_n >
m_\chi + m_{A'}$.  Since $A'$ is unstable because of 
kinetic mixing with the photon, we assume that it is
either sufficiently
long-lived to escape detection, or its decay products $e^+e^-$ or
$3\gamma$ have not yet been excluded by observations of neutron decay.

The decay rate for $n\to \chi A'$ can be expressed in terms of the
ratios $x = (m_n-m_\chi)/m_n$ and $r = m_{A'}/(m_\chi-m_n)$, where
$x \in [0,\,1.78\times 10^{-3}]$ (the upper limit is imposed by the stability of 
$^9$Be \cite{McKeen:2015cuz,Fornal:2018eol}) and $r\in [0,1]$:
\bea
	\Gamma(n\to\chi A') &\cong & {g'^2\theta^2 m_n x\over 2\pi r^2}
	\,(1-r^2)^{3/2}\nn\\
	&\cong& {g'^2 (\delta m)^2 m_n\over 8\pi m_{A'}^2}
	\,x\, (1-r^2)^{3/2}
\eea
where we have ignored negligible higher order corrections in $x$,
and in the second line eliminated $\theta$ using (\ref{mixing}).
The limit $m_{A'}\to 0$ is not singular, despite appearances, when
$A'$ gets its mass from spontaneous symmetry breaking, as we 
explicitly show in section \ref{UV_sec}.  $g' \delta m/ m_{A'}$ remains finite in this
limit.

 To resolve the
neutron lifetime discrepancy, the partial width must be $7\times
10^{-30}$\,GeV \cite{Fornal:2018eol}. For definiteness, we 
adopt the reference value
$m_\chi = 937.9$\,MeV, which maximizes the dark decay rate with
$m_n - m_\chi \cong 1.67$\,MeV.
Moreover we  consider two
benchmark values of $m_{A'} = 1.35$\,MeV and 0.5\,MeV to illustrate
the differences between being above or below the $2 m_e$ threshold for
$A'$ decays. We refer to the two cases as scenarios {\bf A} and
{\bf B}, respectively.  We then find
\be
	{g'\delta m\over m_{A'} } \cong 7.2\,(3.5)\times
10^{-13},\quad \hbox{scenario {\textbf{A }(\textbf{B})}}
\label{mixdet}
\ee
With the neutron star bound $m_{A'}/g'\lesssim 60\,$MeV to be derived 
in section \ref{NS_sec}, this implies
$\delta m < 4\,(2)\times 10^{-11}$\,MeV, which gives a limit on the mixing
angle
\be
	\theta < 1.3\times 10^{-11} \, (6.3\times 10^{-12})
\label{theta_lim}
\ee
for the two scenarios. 
Although the decay $\chi \to p e^- \bar \nu_e$ is kinematically
allowed if $m_\chi > m_p + m_{e^-} = 938.8$ MeV, such a small mixing
endows $\chi$ with a lifetime well beyond the age of the universe.
Therefore $\chi$ is always a component of the DM, even when it is not
absolutely stable.

To satisfy the constraint on $n\to\chi\gamma$ from ref.\ 
\cite{Tang:2018eln}, $\delta m$ must be smaller by a factor of 
approximately 2.6 than the value needed for explaining the lifetime
discrepancy in terms of this decay.\footnote{Ref.\ \cite{Tang:2018eln}
does not quite exclude the model with minimum $m_\chi = 937.9$
that we consider, which would lead to a photon of energy 1.66\,MeV.
The maximum photon energy excluded in \cite{Tang:2018eln} is 1.62 MeV.
We assume that this narrow window will be closed by a wider 
search.}
Using the value of $\delta m$ determined in \cite{Fornal:2018eol}, this leads to the limit
\be
	\delta m \lesssim 3.6\times 10^{-11}\,{\rm MeV}
\label{dmlimit}
\ee
which is similar to those coming from 
(\ref{mixdet}) combined with the bound on $m_{A'}/g'$
from neutron stars, that we  discuss in the next section. This result can also be translated into to a limit on $g'$ using Eqn.~\ref{mixdet}:
\be
	g' \gtrsim 27\,(4.9)\times 10^{-3}, \quad \hbox{scenario {\textbf{A} (\textbf{B})}}
\ee

%---------------------------------------------------------------------------------------------------
%
\begin{figure}[t]
%\hspace{-0.4cm}
\centerline{
\includegraphics[width=\hsize]{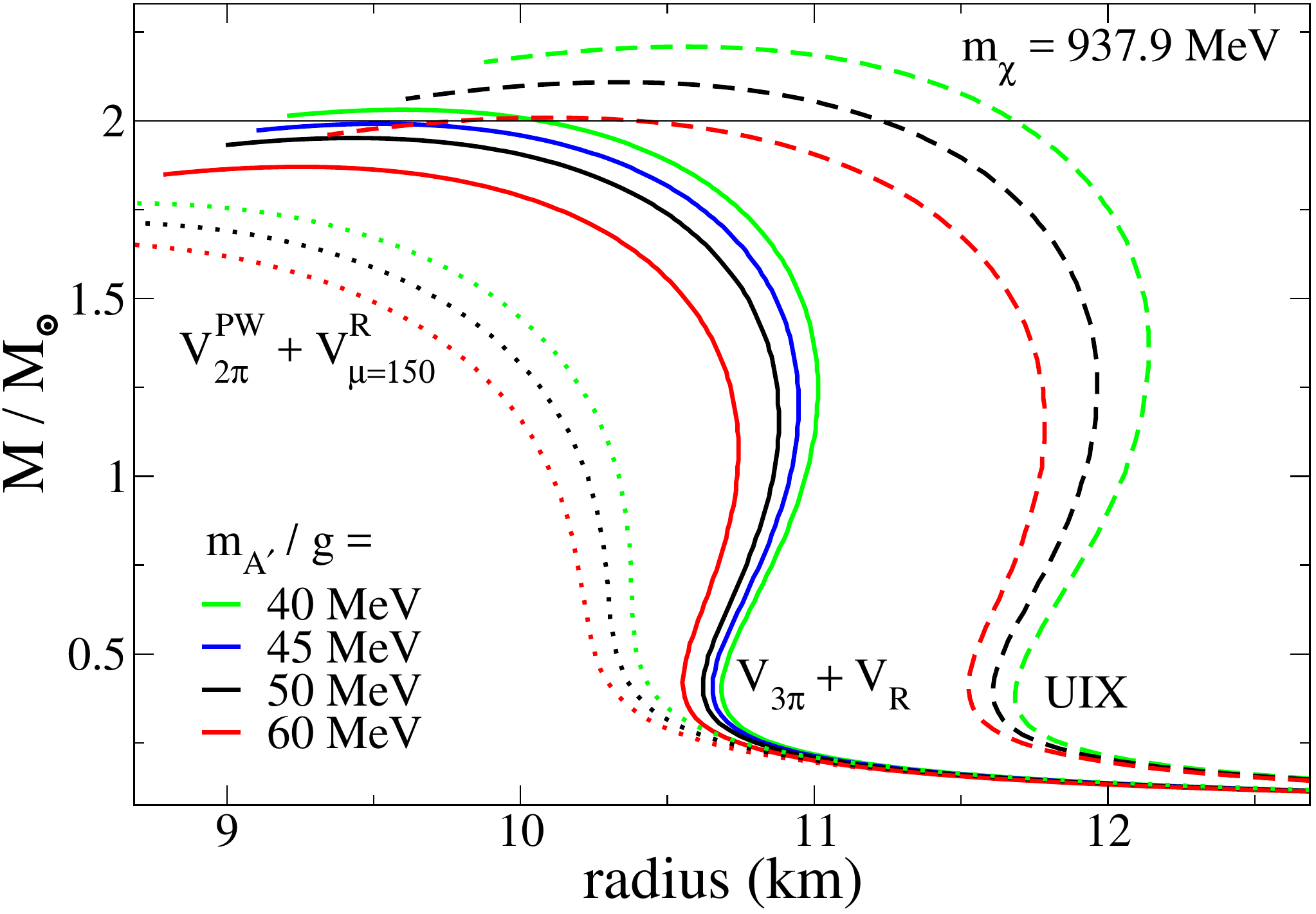}}
\caption{Neutron star mass-radius curves showing the effect of dark
matter pressure for several values of the dark photon mass to charge
ratio $m_{A'}/g'$, and several nuclear equations of state; see text
for details.}
\label{mrfig}
\end{figure}
%
%---------------------------------------------------------------------------------------------------

\section{Neutron star constraints}
\label{NS_sec}
  Although the mixing angle 
(\ref{mixing}) between the
neutron and $\chi$ is very small, it is sufficient to bring the
two into chemical equilibrium in neutron stars \cite{McKeen:2018xwc}.
If $\chi$ is noninteracting, then dense neutron matter 
converts to $\chi$ in order to lower the pressure.  Integrating the
Tolman-Oppenheimer-Volkoff equations \cite{Tolman:1939jz,Oppenheimer:1939ne}
that determine the NS structure,
one finds that the maximum NS mass attained as a function of its radius
falls below the largest observed masses, close to $2 M_\odot$
\cite{2010Natur.467.1081D}.
Ref.\ \cite{McKeen:2018xwc} notes that this can be avoided if 
$m_\chi > 1.2$\,GeV, which would preclude the $n\to\chi$ decays,
or if the DM has repulsive self-interactions, such as would arise from Coulomb
repulsion if $\chi$ is charged under a U(1)$'$ gauge symmetry.  If the
U(1)$'$ is spontaneously broken as we assume, so the gauge boson
is massive, this reduces the pressure since the range of the
interaction becomes $m_{A'}^{-1}$.  We would like to know how large
$m_{A'}/g'$ can be while still attaining a 2 solar mass neutron star.

We have followed the procedure outlined in \cite{McKeen:2018xwc} to
answer this question.  The DM is treated as a degenerate Fermi gas
with self-interactions, so that its energy density is given by
\be
	\rho_\chi = {1\over\pi^2}\int_0^{k_F}dk\, k^2\sqrt{k^2+m_\chi^2}
	+ {g'^2\over 2m_{A'}^2} n_\chi^2
\ee
where $k_F$ is the Fermi momentum and $n_\chi = k_F^3/3\pi^2$.
$k_F$ is determined by the chemical potential
\be
	\mu_\chi = \sqrt{k_F^2+m_\chi^2} + {g'^2\over m_{A'}^2} 
	n_\chi
\ee
The DM pressure is given by $p_\chi = -\rho_\chi + \mu_\chi n_\chi$.
Chemical equilibrium implies that $\mu_\chi$ equals the baryon
(neutron) chemical potential $\mu_B$, which is determined by the
nuclear equation of state (EoS).  $\chi$ will be produced at any 
radius within the NS where $\mu_B > m_\chi$.  The DM contributions 
to pressure and energy density are added to those of the nuclear
matter to find the modified EoS.  

The nuclear EoS can be specified in terms of
the internal energy of a nucleon, $E_n(x)$, as a function of the
dimensionless nuclear density $x = n_n/n_0$, where $n_0 =
0.16$\,fm$^{-3}$ is the saturation density.  We adopt a polytropic
EoS of the form \cite{Gandolfi:2011xu}
\be
	E_n(x) = a x^\alpha + b x^\beta
\ee
that has been fit to the results of quantum Monte Carlo calculations
incorporating realistic 3-nucleon forces.
The nuclear pressure is given by $p_n = n_n^2 dE_n/dn = n_0(a\alpha x^\alpha + 
b\beta x^\beta)$, and the energy density is $\rho_n = n_n(m_n + E_n)$.
The chemical potential is then $\mu_B = (p_n+\rho_n)/n_n$.  
This approximation is adequate at high densities near the center of
the NS, but at some radius as the density increases, the effect of
nuclear binding becomes important, allowing $\mu_B$ to fall below
$m_n$ (which otherwise would not happen) and also $m_\chi$.  At this
radius the $\chi$ density falls to zero and the dark matter can no
longer soften the EoS.  We model this effect by subtracting the
binding energy $E_b = 8$\,MeV (which follows from the semi-empirical
mass formula of nuclear physics  \cite{sanjay}) from $\mu_B$.

Carrying out the above procedure we find the NS mass as a function of 
its radius for a range of central densities of the NS,  to produce
the  mass-radius curves corresponding to different values of
$m_{A'}/g'$.  The results are shown in fig.\ \ref{mrfig} for several
equations of state from ref.\ \cite{Gandolfi:2011xu}.  The solid
lines (labeled $V_{3\pi}+V_R$) correspond to a moderately stiff
EoS  with $a\, (b) = 13.0\, (3.21)$\,MeV,
$\alpha\, (\beta) = 0.49\, (2.47)$, while the dashed ones (labeled
UIX) pertain to a more stiff EoS with 
$a\, (b) = 13.4\, (5.62)$\,MeV,
$\alpha\, (\beta) = 0.514\, (2.436)$.  The dotted curves (labeled
$V_{2\pi}^{PW}+V^R_{\mu=150}$) are for a soft EoS that cannot produce
a 2$M_\odot$ NS even in the absence of dark matter.  
 To obtain a maximum mass compatible
with $2 M_\odot$ requires 
\be
	{m_{A'}\over g'} \lesssim (45-60)\, {\rm MeV}
\label{NSlimit}
\ee
depending upon the nuclear EoS.

\section{\bf UV model}
\label{UV_sec}
Direct and indirect signals of subdominant $\chi$ dark matter can
depend upon details of the particle physics beyond the low-energy
effective description (\ref{Leff}).  In particular a light scalar
particle must be in the spectrum, to account for the mass of $A'$
through the Higgs mechanism.  Here we build a minimal model
that could consistently give rise to the Lagrangian (\ref{Leff}).
It is strongly constrained by the gauge symmetries, giving us little
freedom in its construction.  We demand that the U(1)$'$
gauge symmetry be only spontaneously broken, requiring a scalar $\phi$
that carries U(1)$'$ charge, with a  potential
\be
	V = \lambda\left(|\phi|^2 - v'^2\right)^2	\ .
\ee
Moreover a renormalizable interaction
of $\chi$ with a quark needs a scalar $\Phi_1$ that is fundamental
under SU(3)$_c$ and also charged under U(1)$'$.  To connect this
to the other quarks in the neutron, a second triplet scalar $\Phi_2$,
neutral under U(1)$'$ is needed, along with a cubic scalar interaction,
\be
	\mu\,\Phi_{1,a}\, \Phi_{2}^{*a}\,\phi
\label{cubic}
\ee
where $\mu$ has dimensions of mass.  To satisfy SU(2)$_L$ gauge
symmetry we take the triplets to couple only to 
right-handed quarks,
\be
	\lambda_1 \,\bar d^a P_L \chi\,\Phi_{1,a}
+\lambda_2\, \epsilon^{abc}\,\bar u^c_a P_R\, d_b\, \Phi_{2,c}\,.
\ee
$\Phi_1^*$, $\chi$ and $\phi$ all carry unit dark $U(1)'$ charge with
coupling $g'$.  The vacuum expectation value (VEV) of $\phi$ breaks $U(1)'$ and gives the dark
photon mass $m_{A'} = g'v'$.

We can consistently assign baryon number to all the new fields:
$B_\chi = 1$, $B_{\Phi_1} = B_{\Phi_2} = -2/3$, $B_\phi = 0$.
Therefore the model is free from constraints on $B$-violating
processes such as neutron-antineutron oscillations and dinucleon decay. 
The need to kinematically forbid decays of $^9$Be (if $m_\chi \geq 937.9$
MeV) also stabilizes the proton and forbids possibly problematic
decays of $\Lambda$ baryons to final states including a $\chi$, 
such as
$\Lambda_0 \to \chi \bar K_0$, as well as decays of light mesons to
$\bar \chi \chi$. 

The virtual exchange of the heavy $\Phi_2$
generates a contact interaction that can be Fierz-transformed to \be
{\cal L}_c = {|\lambda_2|^2\over 2 m_{\Phi_2}^2 } (\bar u_\R
\gamma_\mu u_\R) (\bar d_\R \gamma^\mu d_\R) \ee The coefficient is
constrained by measurements of dijet angular distributions at the LHC
\cite{Sirunyan:2017ygf},  $m_{\Phi_2}/\lambda_2 \gtrsim 4\,$TeV.

The scalar triplets, which could decay to two jets or a jet and
missing energy, 
must be at the TeV scale to satisfy LHC constraints on their 
direct production.
\cite{Aaboud:2017vwy}.  Integrating them out generates the
effective interaction
\be
	\eta\, \bar n \phi^* P_L \chi 
\ee
with
\be
	\eta = {\beta \mu \lambda_1 \lambda_2 \over 
	 m^2_{\Phi_1} m^2_{\Phi_2} } \cong
2.4\times 10^{-12}\left(\lambda_1\lambda_2\,\mu\over
	{\rm TeV}\right)
\label{etalim}
\ee
where $\beta \cong  0.014$\,GeV$^3$ from lattice QCD 
\cite{Aoki:2017puj}, and the scalar masses saturate the ATLAS limit 
$m_{\Phi_i} >1.55$\,TeV.
When $\phi$
gets its VEV, the off-diagonal mass $\delta m =
\eta v'$ between $n_R$ and $\chi_L$
is generated.    Since $m_{A'}/g'=v'$, 
we can combine (\ref{etalim}) with (\ref{mixdet}) to determine
\be
	\lambda_1\lambda_2\,\mu \cong (140-290)\,{\rm GeV}
\label{l1l2mu}
\ee
If $\lambda_1\sim\lambda_2\sim 0.4$ (consistent with the previous
LHC limits), then $\mu$ can be near the TeV
scale, like the triplet scalar masses.  But $v'$ must be
at the much lower scale $\lesssim 60$\,MeV from (\ref{NSlimit}).

The scalar $\phi$ cannot be arbitrarily heavy.  Its mass is
$m_\phi = 2\sqrt{\lambda}v$ hence $m_\phi/m_{A'} =
2\sqrt{\lambda}/g'$.  The upper limit from partial wave unitarity is
$\lambda \le 4\pi$.   For our benchmark models this would give a mass
of order $m_\phi \sim 70\,$MeV.  
The decay channel $n\to\chi\phi$ is closed, and only leads to a
subdominant mode $n\to\chi A'A'$ through virtual $\phi$ exchange.
However, when $m_{\phi} \gg m_{A'}$, the scalar decays
via $\phi\to A'A'$, and as $\chi \bar \chi \to \phi A'$ is an allowed process, it can have non-negligible effects on both cosmology and indirect detection signals.
For simplicity we  assume that $\lambda$ is large and $m_\phi$ takes
the benchmark value of 70 MeV. This is large enough so that we can ignore the
effects of $\phi$ on neutron decay, but we will consider its effect on DM physics. We also note the cubic coupling (\ref{cubic})
could lead to interesting cascade decays of the heavier of the two
color triplet scalars $\Phi_i$ produced at the LHC, but this is beyond the
scope of the present work.

\section{Cosmology} 
\label{cosmo_sec}
In the limit of low relative velocity, the cross section for $\chi\chi$ elastic 
scattering by dark photon exchange takes the form
\be
	\sigma_{\chi\chi} = {g'^4 m_\chi^2\over 4 \pi m_{A'}^4}
	\gtrsim 2\times 10^{-24}\,{\rm cm}^2 \, ,
\ee
with the lower bound derived using the limit on $m_{A'}/g'$ from
neutron stars in Eqn.~\ref{NSlimit}. It is interesting that
the lowest allowed value of $\sigma_{\chi\chi}$ is of the appropriate
strength to explain observations of astrophysical structure on small
scales that are in tension with the predictions of cold dark matter
\cite{Tulin:2017ara}.   However constraints from the Cosmic Microwave
Background (CMB) and astrophysics discussed in Section
\ref{ind_det_sec} will ultimately limit $\chi$ to be a small fraction
of the total DM, preventing our model from addressing these
issues.\footnote{For a more complicated dark sector model which could
explain both the neutron lifetime puzzle and small scale
structure issues, see \cite{Karananas:2018goc}.}  Larger values of
$\sigma_{\chi\chi}$ could run afoul of
constraints on DM self-scattering from the Bullet Cluster
\cite{Randall:2007ph}, but this limit is avoided when $\chi$ is
subdominant; up to around 10\% of the total DM population might have
such strong self-interactions  \cite{Kaplan:2009de,Pollack:2014rja}.

The relic density of $\chi$ is determined by $\chi\bar\chi\to A'A'$ and 
$\chi\bar\chi\to \phi A'$ 
annihilations, assuming there is no asymmetry between $\chi$ and
$\bar\chi$. There are also a tree level process that give annihilation into quarks, $\chi \bar \chi \to d\bar d$,
but this process is
greatly suppressed by the TeV scale mass of the $\Phi_2$. The cross sections for the relevant processes are given by \cite{Bell:2016uhg}
\begin{widetext}
\begin{align}
	\label{eq:xsec}
	\langle \sigma_{\rm ann}v_{\rm rel} \rangle_{\chi\bar\chi\to A'A'} &= 
	{g'^4\over 16\pi m_\chi^2}\,
	{(1 - \eta_{A'})^{3/2}\over (1-\eta_{A'}/2)^2}\\
	\label{eq:xsecPhi}		
	\langle \sigma_{\rm ann}v_{\rm rel} \rangle_{\chi\bar\chi\to \phi A'} &=
	\frac{g'^4}{4096 \pi m_\chi^2}
	\frac{\sqrt{\left(\eta_\phi-\eta_{A'} - 4 \right)^2 - 16 \eta_{A'}}
	\left(\left(\eta_\phi-\eta_{A'} - 4 \right)^2 + 32 \eta_{A'}^2 \right)}
	{\left(1-\eta_{A'}/4 \right)^2} \, ,
\end{align}
\end{widetext}
where $\eta_{A',\phi} = m_{A',\phi}^2/m_\chi^2$. 
For $\chi$ to make up all of the DM, the total annihilation cross section should
be close to $10^{-25}$cm$^3$/s \cite{Steigman:2012nb}.
If $m_{A'}\ll m_\chi$, $m_\phi = 70$~MeV, and $\chi$ constitutes less than 10\% of the
total DM, we need a 10 times larger cross section, leading to a lower bound of 
$g' > 0.041$.  If this bound
were saturated, then the NS limit (\ref{NSlimit}) 
would read $m_{A'} < 2.5\,$MeV.  For a strong coupling $g'=1$,
the $\chi$ relic density would be $3\times 10^{-7}$ of the total DM density.

Since $\chi$ is a Dirac particle, it is possible that it has an asymmetry from some unknown
mechanism operating in the early universe.  Then it would be
possible to have large $g'$ and a more significant fraction of $\chi$
being the dark matter. Moreover, if the asymmetric component number density happens
to be close to the symmetric contribution,\footnote{If $n_\chi > n_{\bar \chi}$, we define $n_\chi = n_S + n_A$ and $n_{\bar \chi} = n_S - n_A$, where $n_S$ and $n_A$ are the number densities of the symmetric and anti-symmetric components respectively.} then one CP state of $\chi$
(either $\chi$ or its antiparticle) will have a suppressed density, which 
could weaken indirect detection constraints.  We do not pursue this
loophole in the present work, but it should be kept in mind.

$\chi$ and $A'$ chemically decouple at a dark sector temperature of
$m_\chi/30 \approx 31$ MeV.  At this time
the $A'$ are still
relativistic with a large number density, which can lead to several
problems: if they decay during the time of Big Bang Nucleosynthesis
(BBN) their energetic decay products can disturb the production of light nuclei by diluting the baryon-photon ratio as well as causing photodissociation of the nuclei, and decays during
and after recombination can distort the CMB temperature fluctuations.
Moreover when the dark photons become non-relativistic, an early
period of matter-domination can occur which would change the expansion rate of the universe and cause further
problems for BBN.

To avoid these issues, we follow \cite{Cirelli:2016rnw} and require
the dark photons to decay before they exceed half the energy density
of the universe. The temperature of the SM photon bath at the time this
occurs was shown to be
\be
	T_{\rm dom} \approx \frac{4 m_{A'} Y_{A'}}{3f}
\ee
where $Y_{A'} = n_{A'}/s_{SM}$ at $\chi$ freeze-out and $f=1/2$ is the fraction of the universe's energy density made up of dark photons.
$Y_{A'}$ is given by
\be
	Y_{A'} = \frac{45 \zeta(3)}{2 \pi^4} 
\frac{\tilde g_D}{\tilde g_{SM}} \, ,
\ee
where $\tilde g_D / \tilde g_{SM}$ is the ratio of relativistic degrees of 
freedom in the dark sector to that of the SM sector when the two sectors 
thermally decouple. As the regions of parameter space we are interested in lead 
to small scattering cross sections between dark and SM particles, we assume that 
the two sectors thermally decouple when all species are relativistic, so $\tilde 
g_D = 8$, $\tilde g_{SM} = 106.75$, and $Y_{A'} \approx 0.02$. We then require the 
lifetime $\tau_{A'}$  to be less than $H^{-1}(T_{\rm dom})$, the inverse Hubble 
rate at this temperature. This leads to constraints on the kinetic mixing 
parameter $\epsilon$ that we  present in section \ref{eps_sec}. We emphasize that 
the method we have used here to estimate the cosmological constraints on the dark 
photon lifetime is rough, and more detailed studies similar to those undertaken in 
\cite{Scherrer:1987rr,Fradette:2014sza,Berger:2016vxi,Hufnagel:2017dgo} are needed to determine 
this limit with greater precision.

Finally, DM annihilation close to recombination can also
lead to distortions of the CMB anisotropies. Based on 
searches for this effect in temperature and polarization data,
the \textit{Planck} collaboration has determined the following 95\%
CL constraint \cite{Ade:2015xua}:
\be
	f_{\rm eff} \left( \frac{\Omega_\chi}{\Omega_{DM}} \right)^2 \frac{\langle \sigma_{\rm ann} v_{\rm rel} \rangle}{m_\chi} < 8.2 \times 10^{-28} \ {\rm cm^3/s/GeV} \,
\ee
where $f_{\rm eff}$ is an efficiency parameter that depends on the spectrum of 
injected electrons and photons, while $\Omega_{DM}$ and $\Omega_\chi$ denote the 
ratio of the energy densities of the total dark matter and $\chi$, respectively, 
to the critical density. For $m_\chi = 937.9$ MeV, $f_{\rm eff}$ is approximately 
0.5 for DM annihilations to $2e^+e-$ or $3e^+e-$ final states, and 0.4 for 
annihilation to $6\gamma$ or $9\gamma$ final states \cite{Elor:2015bho}. 
Therefore $(\Omega_\chi / \Omega_{DM})^2 \langle \sigma_{\rm ann} v_{\rm rel} 
\rangle \lesssim 1.5 \times 10^{-27}$ cm$^3$/s for scenario \textbf{A} ($A'$ 
decays to $e^+e^-$), while $(\Omega_\chi / \Omega_{DM})^2 \langle \sigma_{\rm 
ann} v_{\rm rel} \rangle \lesssim 1.9 \times 10^{-27}$ cm$^3$/s for scenario 
\textbf{B} ($A'$ decays to $3 \gamma$).

\section{Direct detection}
\label{DD_sect}  
Because of kinetic mixing, $A'$ couples to
standard model (SM) particles of charge $e$ with strength $\epsilon e$.
The cross section for $\chi$ to scatter on protons is
\be
	\sigma_{\chi p} = {4\alpha(g'\epsilon)^2\, \mu_{p\chi}\over 
m_{A'}^4}
\ee
where $\mu_{p\chi} = 469$\,MeV is the reduced mass of $\chi$ and 
the proton.  The CRESST-III experiment \cite{Petricca:2017zdp} 
sets a limit of $10^{-38}$\,cm$^2$ for scattering on nucleons at
$m_\chi\cong 1$\,GeV, which is relaxed by a factor of $\sim 2$
for our model where only protons scatter. This leads to the constraint
\be
	g' \epsilon < 9.85 \times 10^{-11} \left( \frac{m_{A'}}{\rm MeV} \right)^2 \left(\Omega_{DM}\over\Omega_\chi\right)^{1/2} .
\ee
%We can combine this with 
%the NS limit (\ref{NSlimit}) to find
%\be
%	{\epsilon\over g'}\lesssim 2\times 10^{-7}\left(\Omega_{DM}\over
%	\Omega_\chi\right)^{1/2} .
%\ee
A similar, but weaker bound, arises from $\chi$-electron scattering, 
which has been constrained using XENON100 data in ref.\ \cite{Essig:2017kqs}. 

Mixing of $\chi$ with the neutron leads to scattering mediated by the 
strong interactions of the neutron with nucleons, of order
$\theta^4\,\sigma_{np}$, where $\sigma_{np}\cong 20$\,b is the
low-energy cross section for neutron-proton scattering.  In view of
(\ref{theta_lim}), $\theta^4\,\sigma_{np}< 10^{-67}$\,cm$^2$, far from
observability.  In models with $m_\chi>m_n$, as suggested in 
ref.\ \cite{McKeen:2018xwc} for evading the NS constraint, an
inelastic $\chi N\to n N$ cross section of order 
$\theta^2\,\sigma_{np}\sim 10^{-45}$\,cm$^2$ would arise, with
a highly distinctive signature of large energy deposited in the target
material.  This is still 7 orders of magnitude below current
sensitivity of ordinary elastic scattering, but the large release of 
energy might make it more observable.

In addition to these contributions that only depend upon the
low-energy effective description (\ref{Leff}), exchange of the heavy
colored scalar $\Phi_1$ in the UV model leads to scattering of $\chi$
on down quarks, which after Fierz-transforming comes from the operator
\be
	{\lambda_1^2\over 2
	m_{\Phi_1}^2}(\bar\chi_\L\gamma^\mu\chi_\L)
	(\bar d_\R\gamma_\mu d_\R)
\label{4fermion}
\ee
The vector part of the quark current is simply related to that of
nucleons (counting the number of down quarks in the nucleon).  The 
corresponding cross section is  $\lesssim 10^{-40}$\,cm$^2$, several orders
of magnitude below the current limit.

\section{Indirect detection} 
\label{ind_det_sec}

Indirect DM searches can give strong limits the on annihilation rate of $\chi$ particles.
In particular scenario \textbf{B}, where  $A' \to 3\gamma$, could lead to significant fluxes of gamma-rays from nearby dwarf spheroidal galaxies, and the \textit{Fermi}-LAT collaboration has searched for this excess emission \cite{Ackermann:2015zua}. The
spectrum of photons from a single decay of an $A'$ (in the rest frame
of the $A'$) is given by \cite{Pospelov:2008jk}
\be
	\frac{dN}{dE'} = \frac{16 {E'^3}}{17 {m^4_{A'}}}
		\left[1715 - 6210 \frac{E'}{m_{A'}} + 5838 \left(\frac{E'}{m_{A'}} \right)^2 \right] 
\ee
for $0 \leq E' \leq m_{A'}/2$. The spectrum of photons from the annihilation $\chi \bar\chi \to A' A'$ (in the center of mass frame of the two annihilating particles) is then given by the boost equation 
\be
	\frac{dN}{dE}=\frac{2}{(x_+-x_-)}\int^{E\,x_+}_{E\,x_-}\frac{dE'}{E'}\frac{dN}{dE'},
\ee
where $x_\pm = m_\chi/m_{A'}\pm\sqrt{(m_\chi/m_{A'})^2-1}$. In principle, annihilations of $\chi \bar \chi$ to $\phi A'$ will also have some effect on the shape of the average photon spectrum, but for the regions of parameter space of interest the branching fraction to the $\phi A'$ final state is less than 1/5 (see Eqns.~\ref{eq:xsec} and \ref{eq:xsecPhi}). Therefore we make the approximation that $dN/dE$ is determined solely by annihilations of $\chi \bar \chi$ to $A' A'$.
With this
spectrum we can determine the expected flux 
from a particular galaxy. Using the \texttt{gamLike} code
\cite{Workgroup:2017lvb}, we calculate a likelihood function based on 6 years of \textit{Fermi}-LAT observations of 15 dwarf
spheroidal galaxies. With this likelihood we find that the
quantity $(\Omega_\chi / \Omega_{DM})^2 \langle \sigma_{\rm ann} v_{\rm rel}
\rangle < 2.8 \times 10^{-28}$ cm$^3$/s at 95\% CL for 6-photon final states.

Scenario \textbf{A}, in which the dark photons decay to electron positron pairs, could in theory give a large flux of positrons which could be detected by \textit{Voyager} \cite{Voyager} and the AMS-02 \cite{Accardo:2014lma,Aguilar:2014mma} experiment. However, it was shown in \cite{Boudaud:2016mos}
that significant uncertainties in the model of cosmic-ray propagation make it
difficult to reliably constrain dark matter candidates with masses
less than a GeV via these observations, so we do not consider them further. Limits can also be placed on the $\chi \bar \chi$ annihilation rate in this scenario with \textit{Fermi} observations of dwarf spheroidal galaxies, but these limits are subdominant to those from observations of the CMB, which we discussed in Sec.~\ref{cosmo_sec}.

%---------------------------------------------------------------------------------------------------
%
\begin{figure*}[t]
%\hspace{-0.4cm}
\centerline{
\includegraphics[width=0.5\hsize]{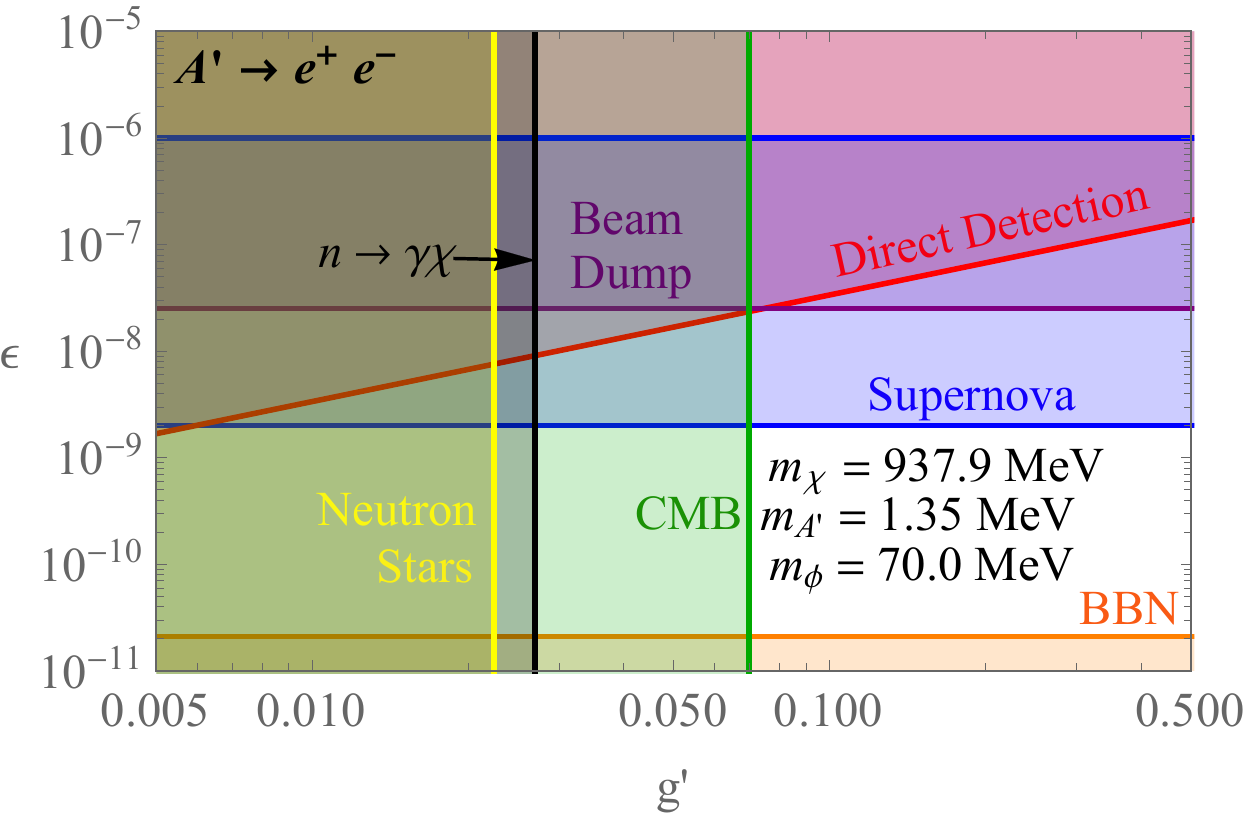}
\includegraphics[width=0.49\hsize]{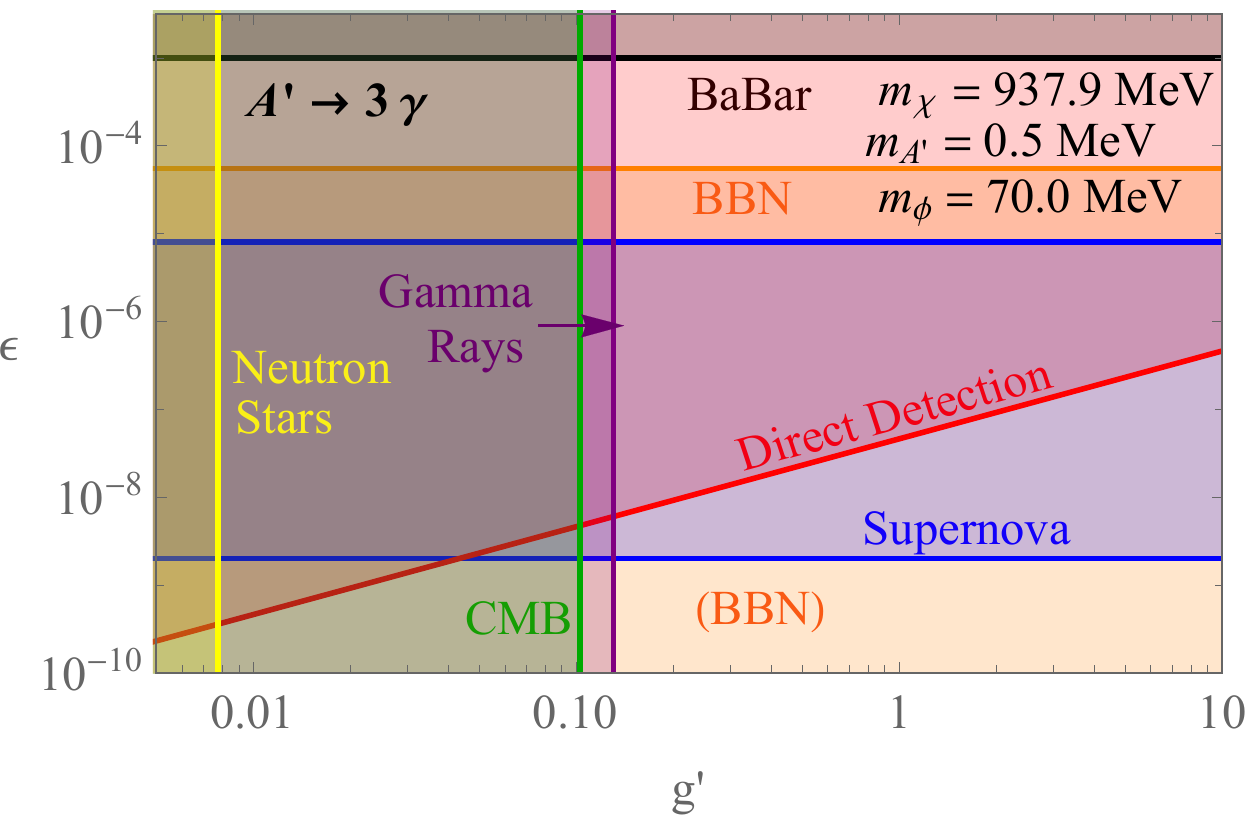}}
\caption{ Limits on $g'$ and $\epsilon$. All shaded regions are excluded. The left plot shows the
constraints on the parameter space when $m_{A'} = 1.35$\,MeV and the
dark photon decays nearly entirely to electron pairs while the right
plot shows limits for when $m_{A'} = 0.5$\,MeV and the predominant
decay is to 3 photons. Constraints shown include those from direct
detection with the CRESST-III experiment \cite{Petricca:2017zdp}
(red; see Sec.\ \ref{DD_sect}), the maximum allowed neutron star mass (yellow; see
Sec.~\ref{NS_sec}), distortions of the CMB anisotropy power spectrum
(green; see
Sec.\ \ref{glimit_sec}) from $\chi$ annihilations \cite{Ade:2015xua}, observations of
the neutrinos from supernova 1987A \cite{Chang:2016ntp} (blue; see
Sec.\ \ref{eps_sec}), BBN
constraints on the decay rate of the dark photons (orange; see
Sec.~\ref{cosmo_sec}), results from the E137 beam dump experiment
\cite{Andreas:2012mt} (purple, left plot; see
Sec.\ \ref{eps_sec}), \textit{BaBar} searches for dark photon production \cite{Lees:2017lec} (black, right plot; see
Sec.\ \ref{eps_sec}), and observations of gamma
rays from dwarf spheroidal galaxies with the \textit{Fermi}-LAT
\cite{Ackermann:2015zua} (purple, right plot; see
Sec.\ \ref{glimit_sec}). The black shaded region in the left plot corresponds to values of $g'$ that cannot explain the neutron lifetime discrepancy due to constraints from searches for neutron decays to a photon and invisible particle \cite{Tang:2018eln} (see Sec.\ \ref{mod_indep_sec}).}
\label{epslimits}
\end{figure*}
%
%---------------------------------------------------------------------------------------------------

\subsection{Constraints on $g'$.}
\label{glimit_sec}
 As $m_{A'} \ll m_\chi$, the DM
annihilation cross section is subject to large Sommerfeld enhancement
at small relative velocities. This leads to an enhanced annihilation
rate in dwarf spheroidal galaxies and at the time of recombination.
The enhanced cross section is given by $\langle \sigma_{\rm ann}
v_{\rm rel} \rangle = S \langle \sigma_{\rm ann} v_{\rm rel}
\rangle_0$, where $\langle \sigma_{\rm ann} v_{\rm rel} \rangle_0$ is
the sum of Eqns.~\ref{eq:xsec} and \ref{eq:xsecPhi} and the Sommerfeld factor is
approximately \cite{Cassel:2009wt} 
\be
	S = \left| \frac{\Gamma(a_-)\Gamma(a_+)}{\Gamma(1 + 2iu)} \right|^2
\ee
with $a_\pm = 1 + iu \left(1 \pm \sqrt{1-x/u} \right)$, $x = {g'}^2/4 \pi \beta$, $u = 6 \beta m_\chi / \pi^2 m_{A'}$, and $\beta = v_{\rm rel}/c$. For all of the dwarf spheroidal galaxies, we make the approximation $\beta \sim 10^{-4}$ (for a compilation of the velocity dispersions in each dwarf spheroidal of interest, see \cite{Choquette:2016xsw}) and we take $\beta \sim 10^{-8}$ at times around recombination. 

Increasing $g'$ ultimately reduces the expected signal in 
searches for DM annihilation, as the increase in the annihilation rate from the increased
cross section is more than offset by the reduced $\chi$ relic density.
For scenario \textbf{A} ($A'$ decays to $e^+ e^-$), the CMB constraints described in Sec.~\ref{cosmo_sec} limit $g' \gtrsim 0.07$, so
$\chi$ can be no more than 1.2\% of the total DM, while for scenario \textbf{B} ($A'$ decays to $3 \gamma$), $g' > 0.11$. For scenario
\textbf{B}, the gamma-ray limits are even stronger, with $g'$
limited to values greater than 0.14. This corresponds to $\chi$ making
up only 0.08\% of the total DM. These limits are displayed in Fig.~\ref{epslimits}. Lowering the mass of the dark photon increases the Sommerfeld enhancement and shifts the limits on $g'$ to higher values, as more suppression of the $\chi$ relic density is needed to avoid these constraints.

\section{Kinetic mixing constraints}
\label{eps_sec}

 The \textit{BaBar} experiment has searched for production of $A'$-photon pairs
in $e+$ $e-$ collisions, with the $A'$ taken to decay to invisible
particles \cite{Lees:2017lec}.  This leads to the constraint $\epsilon <
10^{-3}$. Since the dark photon in our model will decay outside the detector 
when $\epsilon
= 10^{-3}$, this limit applies here. Further direct search
limits come from beam dump experiments, where dark photons with mass
$\lesssim 2\,$MeV can be produced (and detected if $m_{A'} > 2 m_e$). 
Ref.\ \cite{Andreas:2012mt} finds a limit of  $\epsilon < 2.5\times
10^{-8}$ from the E137 experiment.  In addition, if $A'$ is emitted
copiously from supernovae, the observed emission of neutrinos from
supernova 1987A would have been attenuated. This robustly excludes the
overlapping range $\epsilon\in [2\times 10^{-9},\, 1\times 10^{-6}]$
\cite{Chang:2016ntp} (see also \cite{Rrapaj:2015wgs,Hardy:2016kme}).
If $m_{A'}$ is below the $2 m_e$ threshold, the beam dump limits do
not apply, and the supernova limits shift to $\epsilon\in [2\times
10^{-9},\, 8\times 10^{-6}]$.

A lower bound on $\epsilon$ comes from the requirement that $A'$
decays sufficiently fast (before recombination) 
so as not to disturb the CMB. For scenario \textbf{A}, the width of the dominant $A'\to e^+e^-$ decay and corresponding lifetime are given by
\bea
	\Gamma_{e^+e^-} &=& {{\alpha\epsilon^2}\over 3\, m_{A'}}
	\left(m_{A'}^2 + 2 m_e^2\right)\left(1 - 4
m_e^2/m_{A'}^2\right)^{1/2}\nn\\
	\tau_{e^+e^-} &\cong & 24\left(10^{-10}\over\epsilon\right)^2\,{\rm s} \, .
\eea
Using the rate for $A'\to 3\gamma$ 
from \cite{Pospelov:2008jk}, the lifetime for the $A'$ when $m_{A'} < 2 m_e$ is
\be
	\tau_{3\gamma} \cong 2\times 10^{12}\,{\rm s}
      \left(10^{-10}\over\epsilon\right)^2 \left(1\,{\rm MeV}\over
	m_{A'}\right)^9 \, .
\ee

The exact upper limit on $\tau$ from the CMB has not
been worked out in detail in the literature; published constraints on
the allowed abundance of $A'$ versus lifetime only go down to
$\tau = 10^{12}$\,s, where the constraints are rapidly weakening,
but not yet gone; see for example 
fig.\ 7 of ref.\ \cite{Slatyer:2012yq} or fig.\ 4
of ref.\ \cite{Cline:2013fm}.  The true limit is probably somewhat
lower than $10^{12}$\,s; for example if 
$\tau_{3\gamma} < 10^{11}$\,s
and $m_{A'} = 0.5$\,MeV, we require $\epsilon > 4.5\times
10^{-10}$, but since this overlaps with the BBN limits shown in
Fig.\ \ref{epslimits} (right), this uncertainty is not important for
our constraints.

Concerning limits from BBN, following the argument of section
\ref{cosmo_sec}, the requirement for the lifetime of $A'$ is
$\tau_{e^+e^-} < 540$\,s for $m_{A'} = 1.35$\,MeV and $\tau_{3\gamma}
<  3920$\,s for $m_{A'} = 0.5$\,MeV. This translates into lower bounds
on $\epsilon$ of $2 \times 10^{-11}$ and $5 \times 10^{-5}$ for
scenarios \textbf{A} and \textbf{B} respectively, as displayed in
Fig.\ \ref{epslimits} along with the other constraints derived in
previous sections. 
These plots show that the combination of direct detection and
cosmological constraints strongly disfavor $m_{A'} < 1.0$ MeV---only 
when  $A'$ is heavy enough to decay at tree level
are the BBN
constraints on $\epsilon$ sufficiently weak
for the model to be viable.
Lower values of $m_{A'}$ require much larger $\epsilon$
to ensure that $A'$ decays before disturbing BBN.

\section{Conclusions} 

\label{conc_sec}  The possibility that the neutron decays into a dark
fermion $\chi$ and a dark photon $A'$ is highly constrained by an
array of laboratory and astrophysical observations.  The mass of
$\chi$ must lie in a narrow window between $937.9$ and 939.6 MeV. 
The mass of the $A'$ must be greater
than 1.022 MeV ($2 m_e$) and less than 1.67\,MeV.\footnote{In an
extended model, one might circumvent the lower limit on $m_{A'}$
by allowing for $\chi$ to be a long-lived unstable particle with
invisible decay products, perhaps including the true dark matter
of the universe.  Such a framework might be able to circumvent the direct 
detection constraints that lead to $m_{A'} < 2 m_e$ being excluded
in our model.}
 The U(1)$'$ gauge coupling
cannot be small (above 0.07), implying that $\chi$ comprises no more
than $\sim 1$\% of  the total DM.  The kinetic mixing parameter must
lie in a rather narrow range of $10^{-11}$ -- $10^{-9}$.  It is
intriguing that despite the many restrictions, there is still viable
parameter space.   

The decay mode $n\to\chi\gamma$ is still present in our model, but
its rate is suppressed relative to the invisible mode by
$\Gamma(n\to\chi\gamma)/\Gamma(n\to\chi A')\sim(\mu_n v')^2$
where $\mu_n$ is the neutron magnetic moment and $v'$ is the VEV
of the light scalar $\phi$ that breaks the U(1)$'$ symmetry. 
Since smaller values of $v'$ seem less likely in view of the hierarchy
problem for scalar masses, the eventual detection of the visible decay
mode could be expected. Dark neutron decay could also lead to effects in nuclear processes, with $^{11}$Be decays discussed as a particularly promising candidate for study in \cite{Pfutzner:2018ieu}. The dark matter could be
discovered by direct detection, despite its small relic density. 
The UV completion of our model also requires heavy scalar quarks
that could be accessible at LHC and detectable through their decays
to first generation quarks or a jet plus missing energy.

\bigskip
{\bf Note added.}  After completion of this work, ref.\ 
\cite{Sun:2018yaw} reported new limits on the decay channel
$n\to \chi e^+e^-$.  Our model predicts such events from off-shell
dark photon exchange, but at a very low rate.  We computed the partial
width for our benchmark models, finding $10^{-37}-10^{-38}$\,MeV,
which is negligible. In addition, 
ref.\ \cite{Chang:2018uxx} presented a new precise determination of the neutron axial vector coupling 
$g_A$ using lattice QCD. 
This value of the coupling is lower than the experimental average
determined in \cite{Czarnecki:2018okw}, suggesting
consistency between a dark decay contribution to the neutron lifetime
and $g_A$.

\acknowledgements
We thank S.\ McDermott, D.\ McKeen, J.\ Shelton, S.\ Reddy and D.\ Zhou for
helpful correspondence. Our work is supported by the Natural Sciences
and Engineering Research Council (NSERC) of Canada. 

\bibliographystyle{apsrev}

\begin{thebibliography}{10}

%\cite{Fornal:2018eol}
\bibitem{Mampe:1993an} 
  W.~Mampe, L.~N.~Bondarenko, V.~I.~Morozov, Y.~N.~Panin and A.~I.~Fomin,
  ``Measuring neutron lifetime by storing ultracold neutrons and detecting inelastically scattered neutrons,''
  JETP Lett.\  {\bf 57}, 82 (1993)
  [Pisma Zh.\ Eksp.\ Teor.\ Fiz.\  {\bf 57}, 77 (1993)].
  %%CITATION = JTPLA,57,82;%%

%\cite{Serebrov:2004zf}
\bibitem{Serebrov:2004zf} 
  A.~Serebrov {\it et al.},
  ``Measurement of the neutron lifetime using a gravitational trap and a low-temperature Fomblin coating,''
  Phys.\ Lett.\ B {\bf 605}, 72 (2005)
  doi:10.1016/j.physletb.2004.11.013
  [nucl-ex/0408009].
  %%CITATION = doi:10.1016/j.physletb.2004.11.013;%%
  
%\cite{Pichlmaier:2010zz}
\bibitem{Pichlmaier:2010zz} 
  A.~Pichlmaier, V.~Varlamov, K.~Schreckenbach and P.~Geltenbort,
  ``Neutron lifetime measurement with the UCN trap-in-trap MAMBO II,''
  Phys.\ Lett.\ B {\bf 693}, 221 (2010).
  doi:10.1016/j.physletb.2010.08.032
  %%CITATION = doi:10.1016/j.physletb.2010.08.032;%%
   
%\cite{Steyerl:2012zz}
\bibitem{Steyerl:2012zz} 
  A.~Steyerl, J.~M.~Pendlebury, C.~Kaufman, S.~S.~Malik and A.~M.~Desai,
  ``Quasielastic scattering in the interaction of ultracold neutrons with a liquid wall and application in a reanalysis of the Mambo I neutron-lifetime experiment,''
  Phys.\ Rev.\ C {\bf 85}, 065503 (2012).
  doi:10.1103/PhysRevC.85.065503
  %%CITATION = doi:10.1103/PhysRevC.85.065503;%%

%\cite{Arzumanov:2015tea}
\bibitem{Arzumanov:2015tea} 
  S.~Arzumanov, L.~Bondarenko, S.~Chernyavsky, P.~Geltenbort, V.~Morozov, V.~V.~Nesvizhevsky, Y.~Panin and A.~Strepetov,
  ``A measurement of the neutron lifetime using the method of storage of ultracold neutrons and detection of inelastically up-scattered neutrons,''
  Phys.\ Lett.\ B {\bf 745}, 79 (2015).
  doi:10.1016/j.physletb.2015.04.021
  %%CITATION = doi:10.1016/j.physletb.2015.04.021;%%
  
%\cite{Byrne:1996zz}
\bibitem{Byrne:1996zz} 
  J.~Byrne and P.~G.~Dawber,
  ``A Revised Value for the Neutron Lifetime Measured Using a Penning Trap,''
  Europhys.\ Lett.\  {\bf 33}, 187 (1996).
  doi:10.1209/epl/i1996-00319-x
  %%CITATION = doi:10.1209/epl/i1996-00319-x;%%

%\cite{Yue:2013qrc}
\bibitem{Yue:2013qrc} 
  A.~T.~Yue, M.~S.~Dewey, D.~M.~Gilliam, G.~L.~Greene, A.~B.~Laptev, J.~S.~Nico, W.~M.~Snow and F.~E.~Wietfeldt,
  ``Improved Determination of the Neutron Lifetime,''
  Phys.\ Rev.\ Lett.\  {\bf 111}, no. 22, 222501 (2013)
  doi:10.1103/PhysRevLett.111.222501
  [arXiv:1309.2623 [nucl-ex]].
  %%CITATION = doi:10.1103/PhysRevLett.111.222501;%%

%\cite{Pfutzner:2018ieu}
\bibitem{Fornal:2018eol} 
  B.~Fornal and B.~Grinstein,
  ``Dark Matter Interpretation of the Neutron Decay Anomaly,''
  arXiv:1801.01124 [hep-ph].
  %%CITATION = ARXIV:1801.01124;%%
  %6 citations counted in INSPIRE as of 03 Mar 2018

%\cite{Tang:2018eln}
\bibitem{Tang:2018eln} 
  Z.~Tang {\it et al.},
  ``Search for the Neutron Decay n$\rightarrow$ X+$\gamma$ where X is a dark matter particle,''
  arXiv:1802.01595 [nucl-ex].
  %%CITATION = ARXIV:1802.01595;%%
  %4 citations counted in INSPIRE as of 03 Mar 2018


%\cite{Czarnecki:2018okw}
\bibitem{Czarnecki:2018okw} 
  A.~Czarnecki, W.~J.~Marciano and A.~Sirlin,
  ``The Neutron Lifetime and Axial Coupling Connection,''
  arXiv:1802.01804 [hep-ph].
  %%CITATION = ARXIV:1802.01804;%%
  %3 citations counted in INSPIRE as of 03 Mar 2018

\bibitem{McKeen:2018xwc} 
  D.~McKeen, A.~E.~Nelson, S.~Reddy and D.~Zhou,
  ``Neutron stars exclude light dark baryons,''
  arXiv:1802.08244 [hep-ph].
  %%CITATION = ARXIV:1802.08244;%%

\bibitem{Baym:2018ljz} 
  G.~Baym, D.~H.~Beck, P.~Geltenbort and J.~Shelton,
  ``Coupling neutrons to dark fermions to explain the neutron lifetime anomaly is incompatible with observed neutron stars,''
  arXiv:1802.08282 [hep-ph].
  %%CITATION = ARXIV:1802.08282;%%

%\cite{Motta:2018rxp}
\bibitem{Motta:2018rxp} 
  T.~F.~Motta, P.~A.~M.~Guichon and A.~W.~Thomas,
  ``Implications of Neutron Star Properties for the Existence of Light Dark Matter,''
  arXiv:1802.08427 [nucl-th].
  %%CITATION = ARXIV:1802.08427;%%

%\cite{Aitken:2017wie}
\bibitem{Aitken:2017wie} 
  K.~Aitken, D.~McKeen, T.~Neder and A.~E.~Nelson,
  ``Baryogenesis from Oscillations of Charmed or Beautiful Baryons,''
  Phys.\ Rev.\ D {\bf 96}, no. 7, 075009 (2017)
  doi:10.1103/PhysRevD.96.075009
  [arXiv:1708.01259 [hep-ph]].
  %%CITATION = doi:10.1103/PhysRevD.96.075009;%%
  %3 citations counted in INSPIRE as of 20 Jun 2018

%\cite{McKeen:2015cuz}
\bibitem{McKeen:2015cuz} 
  D.~McKeen and A.~E.~Nelson,
  ``CP Violating Baryon Oscillations,''
  Phys.\ Rev.\ D {\bf 94}, no. 7, 076002 (2016)
  doi:10.1103/PhysRevD.94.076002
  [arXiv:1512.05359 [hep-ph]].
  %%CITATION = doi:10.1103/PhysRevD.94.076002;%%


\bibitem{Tolman:1939jz} 
  R.~C.~Tolman,
  ``Static solutions of Einstein's field equations for spheres of fluid,''
  Phys.\ Rev.\  {\bf 55}, 364 (1939).
  doi:10.1103/PhysRev.55.364
  %%CITATION = doi:10.1103/PhysRev.55.364;%%
  %820 citations counted in INSPIRE as of 04 Mar 2018

%\cite{Oppenheimer:1939ne}
\bibitem{Oppenheimer:1939ne} 
  J.~R.~Oppenheimer and G.~M.~Volkoff,
  ``On Massive neutron cores,''
  Phys.\ Rev.\  {\bf 55}, 374 (1939).
  doi:10.1103/PhysRev.55.374
  %%CITATION = doi:10.1103/PhysRev.55.374;%%
  %1224 citations counted in INSPIRE as of 04 Mar 2018


\bibitem[Demorest et al.(2010)]{2010Natur.467.1081D} Demorest, P.~B., Pennucci, T., Ransom, S.~M., Roberts, M.~S.~E., \& Hessels, J.~W.~T.\ 2010, \nat, 467, 1081 



\bibitem{Gandolfi:2011xu} 
  S.~Gandolfi, J.~Carlson and S.~Reddy,
  ``The maximum mass and radius of neutron stars and the nuclear symmetry energy,''
  Phys.\ Rev.\ C {\bf 85}, 032801 (2012)
  doi:10.1103/PhysRevC.85.032801
  [arXiv:1101.1921 [nucl-th]].
  %%CITATION = doi:10.1103/PhysRevC.85.032801;%%
  %198 citations counted in INSPIRE as of 03 Mar 2018

%\cite{Tolman:1939jz}
\bibitem{sanjay}
S. Reddy, private communication

%\cite{Sirunyan:2017ygf}
\bibitem{Sirunyan:2017ygf} 
  A.~M.~Sirunyan {\it et al.} [CMS Collaboration],
  ``Search for new physics with dijet angular distributions in proton-proton collisions at $ \sqrt{s}=13 $ TeV,''
  JHEP {\bf 1707}, 013 (2017)
  doi:10.1007/JHEP07(2017)013
  [arXiv:1703.09986 [hep-ex]].
  %%CITATION = doi:10.1007/JHEP07(2017)013;%%
  %12 citations counted in INSPIRE as of 24 May 2018

%\cite{Randall:2007ph}
\bibitem{Aaboud:2017vwy} 
  M.~Aaboud {\it et al.} [ATLAS Collaboration],
  ``Search for squarks and gluinos in final states with jets and missing transverse momentum using 36 fb$^{-1}$ of $\sqrt{s}$=13 TeV $pp$ collision data with the ATLAS detector,''
  arXiv:1712.02332 [hep-ex].
  %%CITATION = ARXIV:1712.02332;%%
  %8 citations counted in INSPIRE as of 08 Mar 2018


%\cite{Aoki:2017puj}
\bibitem{Aoki:2017puj} 
  Y.~Aoki, T.~Izubuchi, E.~Shintani and A.~Soni,
  ``Improved lattice computation of proton decay matrix elements,''
  Phys.\ Rev.\ D {\bf 96}, no. 1, 014506 (2017)
  doi:10.1103/PhysRevD.96.014506
  [arXiv:1705.01338 [hep-lat]].
  %%CITATION = doi:10.1103/PhysRevD.96.014506;%%
  %4 citations counted in INSPIRE as of 04 Mar 2018
%\cite{McKeen:2018xwc,Baym:2018ljz,Motta:2018rxp}

%\cite{Tulin:2017ara}
\bibitem{Tulin:2017ara} 
  S.~Tulin and H.~B.~Yu,
  %``Dark Matter Self-interactions and Small Scale Structure,''
  Phys.\ Rept.\  {\bf 730}, 1 (2018)
  doi:10.1016/j.physrep.2017.11.004
  [arXiv:1705.02358 [hep-ph]].
  %%CITATION = doi:10.1016/j.physrep.2017.11.004;%%

%\cite{Karananas:2018goc}
\bibitem{Karananas:2018goc} 
  G.~K.~Karananas and A.~Kassiteridis,
  %``Small-scale structure from neutron dark decay,''
  arXiv:1805.03656 [hep-ph].
  %%CITATION = ARXIV:1805.03656;%%
  
%\cite{Bell:2016uhg}
\bibitem{Randall:2007ph} 
  S.~W.~Randall, M.~Markevitch, D.~Clowe, A.~H.~Gonzalez and M.~Bradac,
  ``Constraints on the Self-Interaction Cross-Section of Dark Matter from Numerical Simulations of the Merging Galaxy Cluster 1E 0657-56,''
  Astrophys.\ J.\  {\bf 679}, 1173 (2008)
  doi:10.1086/587859
  [arXiv:0704.0261 [astro-ph]].
  %%CITATION = doi:10.1086/587859;%%
  %358 citations counted in INSPIRE as of 04 Mar 2018

%\cite{Kaplan:2009de,Pollack:2014rja}
\bibitem{Kaplan:2009de} 
  D.~E.~Kaplan, G.~Z.~Krnjaic, K.~R.~Rehermann and C.~M.~Wells,
  ``Atomic Dark Matter,''
  JCAP {\bf 1005}, 021 (2010)
  doi:10.1088/1475-7516/2010/05/021
  [arXiv:0909.0753 [hep-ph]].
  %%CITATION = doi:10.1088/1475-7516/2010/05/021;%%
  %139 citations counted in INSPIRE as of 08 Mar 2018

%\cite{Pollack:2014rja}
\bibitem{Pollack:2014rja} 
  J.~Pollack, D.~N.~Spergel and P.~J.~Steinhardt,
  ``Supermassive Black Holes from Ultra-Strongly Self-Interacting Dark Matter,''
  Astrophys.\ J.\  {\bf 804}, no. 2, 131 (2015)
  doi:10.1088/0004-637X/804/2/131
  [arXiv:1501.00017 [astro-ph.CO]].
  %%CITATION = doi:10.1088/0004-637X/804/2/131;%%
  %22 citations counted in INSPIRE as of 08 Mar 2018

%\cite{Steigman:2012nb}
\bibitem{Bell:2016uhg} 
  N.~F.~Bell, Y.~Cai and R.~K.~Leane,
  ``Impact of mass generation for spin-1 mediator simplified models,''
  JCAP {\bf 1701}, no. 01, 039 (2017)
  doi:10.1088/1475-7516/2017/01/039
  [arXiv:1610.03063 [hep-ph]].
  %%CITATION = doi:10.1088/1475-7516/2017/01/039;%%
  
%\cite{Ade:2015xua}
\bibitem{Steigman:2012nb} 
  G.~Steigman, B.~Dasgupta and J.~F.~Beacom,
  ``Precise Relic WIMP Abundance and its Impact on Searches for Dark Matter Annihilation,''
  Phys.\ Rev.\ D {\bf 86}, 023506 (2012)
  doi:10.1103/PhysRevD.86.023506
  [arXiv:1204.3622 [hep-ph]].
  %%CITATION = doi:10.1103/PhysRevD.86.023506;%%
  %297 citations counted in INSPIRE as of 08 Mar 2018

%\cite{Aaboud:2017vwy}
\bibitem{Cirelli:2016rnw} 
  M.~Cirelli, P.~Panci, K.~Petraki, F.~Sala and M.~Taoso,
  ``Dark Matter's secret liaisons: phenomenology of a dark U(1) sector with bound states,''
  JCAP {\bf 1705}, no. 05, 036 (2017)
  doi:10.1088/1475-7516/2017/05/036
  [arXiv:1612.07295 [hep-ph]].
  %%CITATION = doi:10.1088/1475-7516/2017/05/036;%%
  
%\cite{Scherrer:1987rr}
\bibitem{Scherrer:1987rr} 
  R.~J.~Scherrer and M.~S.~Turner,
  ``Primordial Nucleosynthesis with Decaying Particles. 1. Entropy Producing Decays,''
  Astrophys.\ J.\  {\bf 331}, 19 (1988).
  doi:10.1086/166534
  %%CITATION = doi:10.1086/166534;%%
  
%\cite{Fradette:2014sza}
\bibitem{Fradette:2014sza} 
  A.~Fradette, M.~Pospelov, J.~Pradler and A.~Ritz,
  ``Cosmological Constraints on Very Dark Photons,''
  Phys.\ Rev.\ D {\bf 90}, no. 3, 035022 (2014)
  doi:10.1103/PhysRevD.90.035022
  [arXiv:1407.0993 [hep-ph]].
  %%CITATION = doi:10.1103/PhysRevD.90.035022;%%
  
%\cite{Berger:2016vxi}
\bibitem{Berger:2016vxi} 
  J.~Berger, K.~Jedamzik and D.~G.~E.~Walker,
  ``Cosmological Constraints on Decoupled Dark Photons and Dark Higgs,''
  JCAP {\bf 1611}, 032 (2016)
  doi:10.1088/1475-7516/2016/11/032
  [arXiv:1605.07195 [hep-ph]].
  %%CITATION = doi:10.1088/1475-7516/2016/11/032;%%
  
%\cite{Hufnagel:2017dgo}
\bibitem{Hufnagel:2017dgo}
  M.~Hufnagel, K.~Schmidt-Hoberg and S.~Wild,
  %``BBN constraints on MeV-scale dark sectors. Part I. Sterile decays,''
  JCAP {\bf 1802} (2018) 044
  doi:10.1088/1475-7516/2018/02/044
  [arXiv:1712.03972 [hep-ph]].
  %%CITATION = doi:10.1088/1475-7516/2018/02/044;%%
  
\bibitem{Ade:2015xua} 
  P.~A.~R.~Ade {\it et al.} [Planck Collaboration],
  ``Planck 2015 results. XIII. Cosmological parameters,''
  Astron.\ Astrophys.\  {\bf 594}, A13 (2016)
  doi:10.1051/0004-6361/201525830
  [arXiv:1502.01589 [astro-ph.CO]].
  %%CITATION = doi:10.1051/0004-6361/201525830;%%

%\cite{Elor:2015bho}
\bibitem{Elor:2015bho} 
  G.~Elor, N.~L.~Rodd, T.~R.~Slatyer and W.~Xue,
  ``Model-Independent Indirect Detection Constraints on Hidden Sector Dark Matter,''
  JCAP {\bf 1606}, no. 06, 024 (2016)
  doi:10.1088/1475-7516/2016/06/024
  [arXiv:1511.08787 [hep-ph]].
  %%CITATION = doi:10.1088/1475-7516/2016/06/024;%%

%\cite{Petricca:2017zdp}
\bibitem{Petricca:2017zdp} 
  F.~Petricca {\it et al.} [CRESST Collaboration],
  ``First results on low-mass dark matter from the CRESST-III experiment,''
  arXiv:1711.07692 [astro-ph.CO].
  %%CITATION = ARXIV:1711.07692;%%
  %12 citations counted in INSPIRE as of 04 Mar 2018

%\cite{Essig:2017kqs}
\bibitem{Essig:2017kqs} 
  R.~Essig, T.~Volansky and T.~T.~Yu,
  ``New Constraints and Prospects for sub-GeV Dark Matter Scattering off Electrons in Xenon,''
  Phys.\ Rev.\ D {\bf 96}, no. 4, 043017 (2017)
  doi:10.1103/PhysRevD.96.043017
  [arXiv:1703.00910 [hep-ph]].
  %%CITATION = doi:10.1103/PhysRevD.96.043017;%%
  %39 citations counted in INSPIRE as of 04 Mar 2018

%\cite{Boudaud:2016mos}
\bibitem{Ackermann:2015zua}
  M.~Ackermann {\it et al.} [Fermi-LAT Collaboration],
  ``Searching for Dark Matter Annihilation from Milky Way Dwarf Spheroidal Galaxies with Six Years of Fermi Large Area Telescope Data,''
  Phys.\ Rev.\ Lett.\  {\bf 115}, no. 23, 231301 (2015)
  doi:10.1103/PhysRevLett.115.231301
  [arXiv:1503.02641 [astro-ph.HE]].
  %%CITATION = doi:10.1103/PhysRevLett.115.231301;%%

%\cite{Pospelov:2008jk}
\bibitem{Pospelov:2008jk} 
  M.~Pospelov, A.~Ritz and M.~B.~Voloshin,
  ``Bosonic super-WIMPs as keV-scale dark matter,''
  Phys.\ Rev.\ D {\bf 78}, 115012 (2008)
  doi:10.1103/PhysRevD.78.115012
  [arXiv:0807.3279 [hep-ph]].
  %%CITATION = doi:10.1103/PhysRevD.78.115012;%%

%\cite{Workgroup:2017lvb}
\bibitem{Workgroup:2017lvb} 
  T.~Bringmann {\it et al.} [The GAMBIT Dark Matter Workgroup],
  ``DarkBit: A GAMBIT module for computing dark matter observables and likelihoods,''
  Eur.\ Phys.\ J.\ C {\bf 77}, no. 12, 831 (2017)
  doi:10.1140/epjc/s10052-017-5155-4
  [arXiv:1705.07920 [hep-ph]].
  %%CITATION = doi:10.1140/epjc/s10052-017-5155-4;%%
  
\bibitem{Voyager}
  E.C.~Stone {\it et al.},
  ``Voyager 1 Observes Low-Energy Galactic Cosmic Rays in a Region Depleted of Heliospheric Ions,''
  Science. {\bf 341}, 6142 (2013)
  doi:10.1126/science.1236408.

%\cite{Accardo:2014lma}
\bibitem{Accardo:2014lma} 
  L.~Accardo {\it et al.} [AMS Collaboration],
  ``High Statistics Measurement of the Positron Fraction in Primary Cosmic Rays of 0.5-500 GeV with the Alpha Magnetic Spectrometer on the International Space Station,''
  Phys.\ Rev.\ Lett.\  {\bf 113}, 121101 (2014).
  doi:10.1103/PhysRevLett.113.121101
  %%CITATION = doi:10.1103/PhysRevLett.113.121101;%%
  
%\cite{Aguilar:2014mma}
\bibitem{Aguilar:2014mma} 
  M.~Aguilar {\it et al.} [AMS Collaboration],
  ``Electron and Positron Fluxes in Primary Cosmic Rays Measured with the Alpha Magnetic Spectrometer on the International Space Station,''
  Phys.\ Rev.\ Lett.\  {\bf 113}, 121102 (2014).
  doi:10.1103/PhysRevLett.113.121102
  %%CITATION = doi:10.1103/PhysRevLett.113.121102;%%
  
%\cite{Boudaud:2016mos}
\bibitem{Boudaud:2016mos} 
  M.~Boudaud, J.~Lavalle and P.~Salati,
  ``Novel cosmic-ray electron and positron constraints on MeV dark matter particles,''
  Phys.\ Rev.\ Lett.\  {\bf 119}, no. 2, 021103 (2017)
  doi:10.1103/PhysRevLett.119.021103
  [arXiv:1612.07698 [astro-ph.HE]].
  %%CITATION = doi:10.1103/PhysRevLett.119.021103;%%

%\cite{Ackermann:2015zua}
\bibitem{Chang:2016ntp} 
  J.~H.~Chang, R.~Essig and S.~D.~McDermott,
  ``Revisiting Supernova 1987A Constraints on Dark Photons,''
  JHEP {\bf 1701}, 107 (2017)
  doi:10.1007/JHEP01(2017)107
  [arXiv:1611.03864 [hep-ph]].
  %%CITATION = doi:10.1007/JHEP01(2017)107;%%

\bibitem{Andreas:2012mt} 
  S.~Andreas, C.~Niebuhr and A.~Ringwald,
  ``New Limits on Hidden Photons from Past Electron Beam Dumps,''
  Phys.\ Rev.\ D {\bf 86}, 095019 (2012)
  doi:10.1103/PhysRevD.86.095019
  [arXiv:1209.6083 [hep-ph]].
  %%CITATION = doi:10.1103/PhysRevD.86.095019;%%
  %123 citations counted in INSPIRE as of 06 Mar 2018

%\cite{Lees:2017lec}
\bibitem{Lees:2017lec} 
  J.~P.~Lees {\it et al.} [BaBar Collaboration],
  ``Search for Invisible Decays of a Dark Photon Produced in ${e}^{+}{e}^{-}$ Collisions at BaBar,''
  Phys.\ Rev.\ Lett.\  {\bf 119}, no. 13, 131804 (2017)
  doi:10.1103/PhysRevLett.119.131804
  [arXiv:1702.03327 [hep-ex]].
  %%CITATION = doi:10.1103/PhysRevLett.119.131804;%%
  
%\cite{Chang:2016ntp}
\bibitem{Cassel:2009wt} 
  S.~Cassel,
  ``Sommerfeld factor for arbitrary partial wave processes,''
  J.\ Phys.\ G {\bf 37}, 105009 (2010)
  doi:10.1088/0954-3899/37/10/105009
  [arXiv:0903.5307 [hep-ph]].
  %%CITATION = doi:10.1088/0954-3899/37/10/105009;%%

%\cite{Choquette:2016xsw}
\bibitem{Choquette:2016xsw} 
  J.~Choquette, J.~M.~Cline and J.~M.~Cornell,
  ``p-wave Annihilating Dark Matter from a Decaying Predecessor and the Galactic Center Excess,''
  Phys.\ Rev.\ D {\bf 94}, no. 1, 015018 (2016)
  doi:10.1103/PhysRevD.94.015018
  [arXiv:1604.01039 [hep-ph]].
  %%CITATION = doi:10.1103/PhysRevD.94.015018;%%

%\cite{Andreas:2012mt}
\bibitem{Rrapaj:2015wgs} 
  E.~Rrapaj and S.~Reddy,
  ``Nucleon-nucleon bremsstrahlung of dark gauge bosons and revised supernova constraints,''
  Phys.\ Rev.\ C {\bf 94}, no. 4, 045805 (2016)
  doi:10.1103/PhysRevC.94.045805
  [arXiv:1511.09136 [nucl-th]].
  %%CITATION = doi:10.1103/PhysRevC.94.045805;%%
  %17 citations counted in INSPIRE as of 06 Mar 2018

%\cite{Hardy:2016kme}
\bibitem{Hardy:2016kme} 
  E.~Hardy and R.~Lasenby,
  ``Stellar cooling bounds on new light particles: plasma mixing effects,''
  JHEP {\bf 1702}, 033 (2017)
  doi:10.1007/JHEP02(2017)033
  [arXiv:1611.05852 [hep-ph]].
  %%CITATION = doi:10.1007/JHEP02(2017)033;%%

%\cite{Slatyer:2012yq}
\bibitem{Slatyer:2012yq} 
  T.~R.~Slatyer,
  ``Energy Injection And Absorption In The Cosmic Dark Ages,''
  Phys.\ Rev.\ D {\bf 87}, no. 12, 123513 (2013)
  doi:10.1103/PhysRevD.87.123513
  [arXiv:1211.0283 [astro-ph.CO]].
  %%CITATION = doi:10.1103/PhysRevD.87.123513;%%
  %77 citations counted in INSPIRE as of 07 Mar 2018

%\cite{Cline:2013fm}
\bibitem{Cline:2013fm} 
  J.~M.~Cline and P.~Scott,
  ``Dark Matter CMB Constraints and Likelihoods for Poor Particle Physicists,''
  JCAP {\bf 1303}, 044 (2013)
  Erratum: [JCAP {\bf 1305}, E01 (2013)]
  doi:10.1088/1475-7516/2013/03/044, 10.1088/1475-7516/2013/05/E01
  [arXiv:1301.5908 [astro-ph.CO]].
  %%CITATION = doi:10.1088/1475-7516/2013/03/044, 10.1088/1475-7516/2013/05/E01;%%
  %62 citations counted in INSPIRE as of 07 Mar 2018

%\cite{Mampe:1993an}
\bibitem{Pfutzner:2018ieu} 
  M.~Pf\"utzner and K.~Riisager,
  ``On the possibility to observe neutron dark decay in nuclei,''
  arXiv:1803.01334 [nucl-ex].
  %%CITATION = ARXIV:1803.01334;%%

%\cite{Sun:2018yaw}
\bibitem{Sun:2018yaw} 
  X.~Sun {\it et al.} [UCNA Collaboration],
  ``Search for dark matter decay of the free neutron from the UCNA experiment: n $\rightarrow \chi + e^+e^-$,''
  doi:10.1103/PhysRevC.97.052501
  arXiv:1803.10890 [nucl-ex].
  %%CITATION = doi:10.1103/PhysRevC.97.052501;%%
 
%\cite{Chang:2018uxx}
\bibitem{Chang:2018uxx} 
  C.~C.~Chang {\it et al.},
  ``A per-cent-level determination of the nucleon axial coupling from quantum chromodynamics,''
  Nature {\bf 558}, no. 7708, 91 (2018)
  doi:10.1038/s41586-018-0161-8
  [arXiv:1805.12130 [hep-lat]].
  %%CITATION = doi:10.1038/s41586-018-0161-8;%%
  
\end{thebibliography}

\end{document}